\documentclass{article}

\usepackage{arxiv}

\usepackage[utf8]{inputenc} % allow utf-8 input
\usepackage[T1]{fontenc}    % use 8-bit T1 fonts
\usepackage{hyperref}       % hyperlinks
\usepackage{url}            % simple URL typesetting
\usepackage{booktabs}       % professional-quality tables
\usepackage{amsfonts}       % blackboard math symbols
\usepackage{nicefrac}       % compact symbols for 1/2, etc.
\usepackage{microtype}      % microtypography
\usepackage{lipsum}
\usepackage{graphicx}
\usepackage{tcolorbox}
\usepackage{graphicx}
\graphicspath{ {./img/} }

\title{Towards auto-completion on software requirements statements}

\author{
 Carlos Alberto dos Santos \\
  SmArtSE Research Team\\
  Université du Québec à Chicoutimi\\
  555, boulevard de l’Université, G7H 2B1 \\
  Chicoutimi, QC, Canada \\
  \texttt{carlos-alberto.dos-santos1@uqac.ca} \\

   \And
 Fabio Petrillo \\
  SmArtSE Research Team\\
  Université du Québec à Chicoutimi\\
  555, boulevard de l’Université, G7H 2B1 \\
  Chicoutimi, QC, Canada \\
  \texttt{fabio@petrillo.com} \\
}

\begin{document}
\maketitle

\begin{abstract}
As software systems become more complex, modern software development requires more attention to human perspectives, and active participation of development teams in requirements elicitation tasks. In this context, incomplete or ambiguous requirements descriptions do not guide the development of good software products. We hypothesize that the text auto-completion feature improves the quality of the software requirements artifacts. We present the motivation for this study, related works, our approach and future research efforts.
\end{abstract}


\section{Introduction}
\label{sec:introduction}
As the world becomes more complex, software used in several sectors of society need to scale to cope with the changes. The modern software development requires an holistic view about many perspectives such as human, organizational, social, and political issues, in order to build systems able to interact communities \cite{10.1145/2209249.2209268}. Software developers need to gear up for dealing with these topics besides usual technical factors. 
In front of this scenario, new development processes are constantly emerging as a goal to integrate stakeholders and technical teams. Many of these process have agile mindset \cite{beck2001agile} as basis, such as Extreme Programming \cite{xpKent} and Scrum \cite{schwaber_sutherland, sutherland2014scrum}. The usage of these process can help technical teams to keep the \textbf{development focus on the business}, delivering software \textbf{fit for purpose}, and enabling the \textbf{software evolution}.

The focus of development teams should be on delivering high quality software, even when they have to deal with several business factors during the development process. Technical teams can negotiate project scope, budget and even the deadlines, but rarely software quality. \textbf{Software quality is not negotiable} as the software failures can affect essential process and services.

The software industry has defined the basis of quality management in order to improve the quality software products. \textbf{Software should satisfy the user expectations}, performing efficiently and reliably, delivering value on time and according to the budget. The usage of quality management techniques has been improving software products over last 20 years \cite{sommerville2016software}.

Quality management in agile development relies on establishing a quality culture among development teams. Every single developer is responsible for software quality and synchronized actions are set to ensure that quality is maintained \cite{sommerville2016software}. In this scenario, testing efforts are employed at an earlier point in the development process \cite{10.1007/978-3-642-30561-0_19}. \textbf{The testing step needs to be based on user requirements} as a way to discover any defects that may bring system fails to meet the client's requirements \cite{chopra2018software}.

However, incomplete or ambiguous requirements descriptions do not guide the development of good software products. The client participation during the requirements review step is essential to ensure the quality of the requirements statements as it depends on the domain knowledge \cite{FEMMER2017190}. \textbf{The agile process adoption has been encouraging development teams to engage with business area}. Those teams are taking the ownership about requirements refinement meetings, client communication and user stories writing, as a way to be closer to stakeholders. Agile processes bring the practice of user story writing to represent scenarios of use about situations experienced by a system user \cite{sommerville2016software}. Thus, when the user needs are not well written, they can communicate inaccurate messages to the development team, affecting since the coding until the software delivery.

Based on this scenario, the \textbf{ ISO/IEC 29148 standard} \cite{8559686} was set as reference to requirements descriptions. It brings guidance to construct good system requirements, providing attributes and characteristics of requirements, and discussing the requirements engineering during the project life cycle. Other standards, such as ISO/IEC/IEEE 15288 and ISO/IEC/IEEE 12207 are also associate to software engineering steps and they are used as ground to ISO/IEC 29148.

Femmer \textit{et. al.} \cite{FEMMER2017190} have relied on the ISO requirements standardize to set the concept of \textbf{Requirements Smells}. This concept defines concrete symptoms for a quality violation of requirement artifacts. It is based on a subset of language criteria given by the standard ISO/IEC 29148 \cite{8559686}, used for highlight bad requirements descriptions that can guide to potential negative effects on activities on software life cycle. This concept helps researches to define the bases of requirements artifact quality, and others papers \cite{linkToFemmer1, linkToFemmer2, linkToFemmer3, linkToFemmer4} have been employing this technique in order to deal with requirements quality problem.

Requirements quality is a wide subarea of software engineering constantly drawing attention from researches. Some studies have been employing artificial intelligence techniques to deal with software requirements \cite{reml1, reml2, reml3, reml4, reml5}. These papers support elicitation and categorizing tasks using techniques to ensure higher requirements quality level.

\textbf{This positional paper} presents our approach on adoption of text auto-completion feature to assist the software requirements writing. This approach can collaborate with the requirements quality subarea, proposing a new way to produce higher quality software requirements.

\section{Proposed approach}

Our approach intend to use text auto-completion feature to help developers and business analysts to write software requirements. The auto-completion feature powers-up the quality of requirements artifacts suggesting terms to complete phrases, and also removing ambiguity and orthographic errors. Thus, it accelerates the writing process, keeping the writer focus on the business and ensuring a better system behavior comprehension by other stakeholders involved in the project. The box below presents our hypothesis.

\begin{tcolorbox}[colframe=gray!50, coltitle=black, title=Hypothesis]
	The text auto-completion feature improves the quality of the software requirements artifacts.
	\label{box:hypothesis}
\end{tcolorbox}

For that, we propose an exploration work divided in steps to validate our approach. The first step is to look for recent studies related to software requirements quality, in order to know \textbf{which approaches have been used} to address the requirements quality problem and which are the \textbf{research opportunities} in this area. We want also to review the ISO/IEC standards related to software engineering to understand the theoretical elements about software requirements definition.

Secondly, we intend to make a \textbf{grey literature study} to evaluate how industry tackles the requirements quality topic. Through this study, we want to know the needs of software industry and how poor requirements definitions can affect the development life cycle, highlighting examples and study cases.

\begin{figure}[h!]
    \centering
    \includegraphics[width=0.8\textwidth]{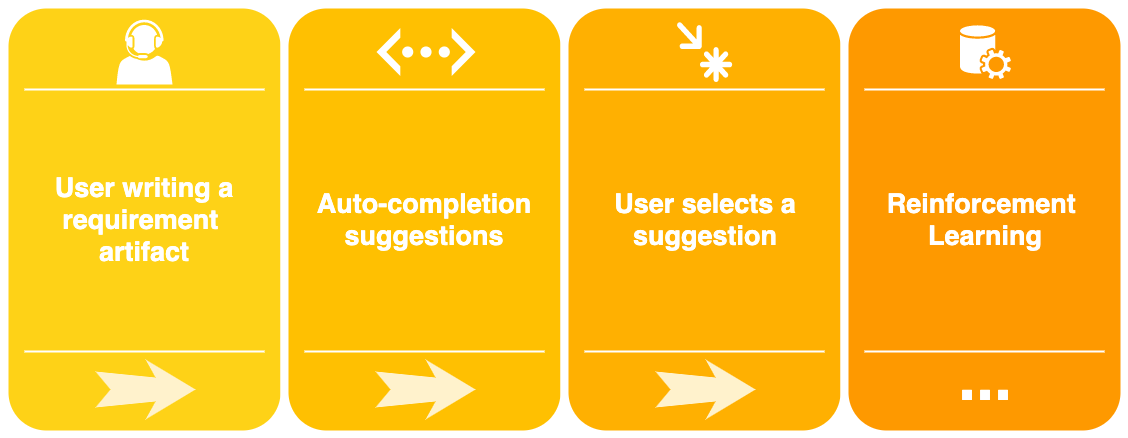}
    \caption{Text auto-completion feature for supporting requirements elicitation.}
    \label{fig:restepts}
\end{figure}

Thus, we want to evaluate how \textbf{requirements support tools} control the text quality, and how they organize the business knowledge in software development projects. These are some examples of tools that support the requirements management: Drools\footnote{https://www.drools.org/}, Cucumber\footnote{https://cucumber.io/} and JBehave\footnote{https://jbehave.org/}.

After the research step, we intend to develop a \textbf{tool prototype} to evaluate our approach with a group of users. All these users will be software developers or business analysts used to write and read software requirements in their projects. We want to assess their experience by using our prototype simulating a requirements definition environment.

This prototype will use \textbf{reinforcement learning} to update the auto-completion possibilities while the user is writing software requirements. The Figure \ref{fig:restepts} shows a workflow about how this tool will work to suggest text auto-completion possibilities to the user.

\section{Related work}

%requirements smells
Femmer \textit{et. al.} \cite{FEMMER2017190} provided a tool called \textit{Smella} in order to automate the requirements smells identification in requirements artifacts. Its implementation is based on part-of-speech (POS) tagging, morphological analysis and dictionaries. This tool has helped the authors to conduce an empirical evaluation in a multi-case study with
different users and prove the efficacy of their approach about Requirements Smells. Their results show that the usage of \textit{Smella} produces a precision of 0.59 and recall of 0.82 in identifying requirements smells in artifacts. Although the low precision demands more checking work, this work is important as it introduces a new approach and presents valuable results referent to requirements quality analysis.

%requirements completeness paper
Ko \textit{et. al.} \cite{Ko2019} produced a tool called \textit{ScenarioAmigo} as a goal to automatically identify completeness in software requirements specifications. Completeness is an important characteristic of software requirements artifacts and it indicates whether the text has no omitted requirements in the context. Incomplete requirements can lead stakeholders to miss important features of systems. The lack of vision about the user scenarios is one of the main causes of incomplete requirements specifications. To deal with this problem, \textit{ScenarioAmigo} was implemented using verb clustering algorithm and scenario flow graphs, in order to detect omitted steps of user’s scenarios, and automatically recommend appropriate steps for the omitted parts. The results present that the comparison of recall of \textit{ScenarioAmigo} to manually requirements revision performed by human experts obtained the 20\% higher score.

Allahyari-Abhari \textit{et. al.} \cite{Allahyari-Abhari} used metrics computed from the text structures to identify syntactic and semantic quality in requirements artifacts, and assist the author in writing good
requirements statements. This paper has used structural parser to generate phrase structure trees (PST), using a \textit{probabilistic context-free grammar} (PCFG) as background for computing the PST values. The waited output is a single quality measure to classify a given sentence in good, medium, or bad. It is based on the criteria that the PST with the highest PCFG score is the
one which is most likely to be correct. This approach is powered-up with machine learning techniques using Naive Bayes classifier. The results reached by the authors show that their approach can automatically classify requirements statements quality with F1-measure of 74.1\%.

In the study performed by Yotaro \textit{et. al.} \cite{8920417} was proposed a technique to automate the bad smells detection in use case descriptions (symptoms of poor descriptions). In this work, the bad smells definition comes from a manually work analyzing use case models, as a goal to discover poor use case descriptions. For implementing their approach was used Mecab, as Part-of-speech parser and NEologd as Japanese dictionary to support the metrics computation to identify bad smells. As a result, the first version of the tool got a precision ratio of 0.591 and recall ratio of 0.981 against manual analysis. The corpus used in this study was created based on 30 use case descriptions of several domains.

\section{Conclusion}

This paper presented our approach about the usage of text auto-completion feature to improve the quality of the software requirements artifacts. In the introduction session, we performed an overview about software requirements engineering, and how modern software development relies on this task to deal with business complexity. Other important aspect described was the standards ISO/IEC useful to guide the software requirements statements writing, and the studies inherited from these standards. As a goal to validate our hypothesis, we defined the following steps: (i) search for recent studies aligned to software requirements quality; (ii) perform a grey literature study to evaluate how the topic is tackled in other outside the academical mainstream; (iii) analysis of requirements support tools; (iv) tool prototype development in order to (v) evaluate our approach using this tool in a requirements definition environment.

Thus, we believe that the text auto-completion feature improves the quality of the software requirements artifacts.

\bibliographystyle{unsrt}  
\bibliography{references}

\begin{thebibliography}{10}

\bibitem{10.1145/2209249.2209268}
Ian Sommerville, Dave Cliff, Radu Calinescu, Justin Keen, Tim Kelly, Marta
  Kwiatkowska, John Mcdermid, and Richard Paige.
\newblock Large-scale complex it systems.
\newblock {\em Commun. ACM}, 55(7):71–77, July 2012.

\bibitem{beck2001agile}
Kent Beck, Mike Beedle, Arie van Bennekum, Alistair Cockburn, Ward Cunningham,
  Martin Fowler, James Grenning, Jim Highsmith, Andrew Hunt, Ron Jeffries, Jon
  Kern, Brian Marick, Robert~C. Martin, Steve Mellor, Ken Schwaber, Jeff
  Sutherland, and Dave Thomas.
\newblock Manifesto for agile software development, 2001.

\bibitem{xpKent}
K.~Beck.
\newblock Embracing change with extreme programming.
\newblock {\em Computer}, 32(10):70--77, 1999.
\newblock Cited By :471.

\bibitem{schwaber_sutherland}
Ken Schwaber and Jeff Sutherland.
\newblock The 2020 scrum guidetm.

\bibitem{sutherland2014scrum}
J.~Sutherland and J.J. Sutherland.
\newblock {\em Scrum: The Art of Doing Twice the Work in Half the Time}.
\newblock Crown, 2014.

\bibitem{sommerville2016software}
I.~Sommerville.
\newblock {\em Software Engineering}.
\newblock Always learning. Pearson, 2016.

\bibitem{10.1007/978-3-642-30561-0_19}
Mathias Soeken, Robert Wille, and Rolf Drechsler.
\newblock Assisted behavior driven development using natural language
  processing.
\newblock In Carlo~A. Furia and Sebastian Nanz, editors, {\em Objects, Models,
  Components, Patterns}, pages 269--287, Berlin, Heidelberg, 2012. Springer
  Berlin Heidelberg.

\bibitem{chopra2018software}
R.~Chopra.
\newblock {\em Software Quality Assurance: A Self-Teaching Introduction}.
\newblock Mercury Learning \& Information, 2018.

\bibitem{FEMMER2017190}
Henning Femmer, Daniel {Méndez Fernández}, Stefan Wagner, and Sebastian Eder.
\newblock Rapid quality assurance with requirements smells.
\newblock {\em Journal of Systems and Software}, 123:190--213, 2017.

\bibitem{8559686}
International~Organization for Standardization~[ISO].
\newblock Iso/iec/ieee international standard - systems and software
  engineering -- life cycle processes -- requirements engineering.
\newblock {\em ISO/IEC/IEEE 29148:2018(E)}, pages 1--104, 2018.

\bibitem{linkToFemmer1}
B.~Rosadini, A.~Ferrari, G.~Gori, A.~Fantechi, S.~Gnesi, I.~Trotta, and
  S.~Bacherini.
\newblock {\em Using NLP to detect requirements defects: An industrial
  experience in the railway domain}, volume 10153 LNCS of {\em Lecture Notes in
  Computer Science (including subseries Lecture Notes in Artificial
  Intelligence and Lecture Notes in Bioinformatics)}.
\newblock 2017.
\newblock Cited By :34.

\bibitem{linkToFemmer2}
A.~Ferrari, G.~Gori, B.~Rosadini, I.~Trotta, S.~Bacherini, A.~Fantechi, and
  S.~Gnesi.
\newblock Detecting requirements defects with nlp patterns: an industrial
  experience in the railway domain.
\newblock {\em Empirical Software Engineering}, 23(6):3684--3733, 2018.
\newblock Cited By :17.

\bibitem{linkToFemmer3}
F.~Dalpiaz, I.~van~der Schalk, S.~Brinkkemper, F.~B. Aydemir, and G.~Lucassen.
\newblock Detecting terminological ambiguity in user stories: Tool and
  experimentation.
\newblock {\em Information and Software Technology}, 110:3--16, 2019.
\newblock Cited By :14.

\bibitem{linkToFemmer4}
A.~Fantechi, A.~Ferrari, S.~Gnesi, and L.~Semini.
\newblock Hacking an ambiguity detection tool to extract variation points: An
  experience report.
\newblock In {\em ACM International Conference Proceeding Series}, pages
  43--50, 2018.
\newblock Cited By :8.

\bibitem{reml1}
S.~Abualhaija, C.~Arora, M.~Sabetzadeh, L.~C. Briand, and M.~Traynor.
\newblock Automated demarcation of requirements in textual specifications: a
  machine learning-based approach.
\newblock {\em Empirical Software Engineering}, 25(6):5454--5497, 2020.
\newblock Cited By :1.

\bibitem{reml2}
Y.~Wang, L.~Shi, M.~Li, Q.~Wang, and Y.~Yang.
\newblock A deep context-wise method for coreference detection in natural
  language requirements.
\newblock In {\em Proceedings of the IEEE International Conference on
  Requirements Engineering}, volume 2020-August, pages 180--191, 2020.

\bibitem{reml3}
K.~Kolthoff.
\newblock Automatic generation of graphical user interface prototypes from
  unrestricted natural language requirements.
\newblock In {\em Proceedings - 2019 34th IEEE/ACM International Conference on
  Automated Software Engineering, ASE 2019}, pages 1234--1237, 2019.
\newblock Cited By :1.

\bibitem{reml4}
F.~Dalpiaz, D.~Dell'Anna, F.~B. Aydemir, and S.~Çevikol.
\newblock Requirements classification with interpretable machine learning and
  dependency parsing.
\newblock In {\em Proceedings of the IEEE International Conference on
  Requirements Engineering}, volume 2019-September, pages 142--152, 2019.
\newblock Cited By :10.

\bibitem{reml5}
D.~Falessi and G.~Cantone.
\newblock The effort savings from using nlp to classify equivalent
  requirements.
\newblock {\em IEEE Software}, 36(1):48--55, 2019.
\newblock Cited By :1.

\bibitem{Ko2019}
Deokyoon Ko, Suntae Kim, and Sooyong Park.
\newblock Automatic recommendation to omitted steps in use case specification.
\newblock {\em Requirements Engineering}, 24(4):431--458, Dec 2019.

\bibitem{Allahyari-Abhari}
A.~Allahyari-Abhari, M.~Soeken, and R.~Drechsler.
\newblock Requirement phrasing assistance using automatic quality assessment.
\newblock In {\em Proceedings - 2015 IEEE 18th International Symposium on
  Design and Diagnostics of Electronic Circuits and Systems, DDECS 2015}, pages
  183--188, 2015.
\newblock Cited By :1.

\bibitem{8920417}
Yotaro Seki, Shinpei Hayashi, and Motoshi Saeki.
\newblock Detecting bad smells in use case descriptions.
\newblock In {\em 2019 IEEE 27th International Requirements Engineering
  Conference (RE)}, pages 98--108, 2019.

\end{thebibliography}

\end{document}